# Giant energy storage effect in nanolayer capacitors charged by the field emission tunneling

Eduard Ilin[1], Irina Burkova[1], Eugene V. Colla[1], Michael Pak[2], and Alexey Bezryadin[1]
[1]Department of Physics, University of Illinois at Urbana-Champaign, Urbana, IL 61801, USA
[2]Department of Engineering Physics, Air Force Institute of Technology, Dayton, OH 45433, USA

**Abstract:** We fabricate nanolayer alumina capacitor and apply high electric fields, close to 1 GV/m, to inject charges in the dielectric. Asymmetric charge distributions have been achieved due to the selectivity of the quantum tunneling process. Namely, the electrons cannot tunnel to a region near cathode, where the total energy would be less than the potential energy. This mechanism exhibits a strong tendency to populate charge traps located near the anode, i.e., the regions where their potential energy is the lowest. This charge injection allows a permanent storage of the bulk charge even if the capacitor plates are short-circuited, provided that the temperature is sufficiently low so that the conductivity of the dielectric is negligible. In our experiments, the total charge stored in the dielectric was up to seven and a half times higher than the charge stored on the capacitor plates. Also, measurements of the breakdown voltage show that the breakdown electric field, i.e., the dielectric strength, is independent of the thickness of the dielectric.

## 1. INTRODUCTION

The energy storage problem is of great importance now since the continuous usage of traditional energy carriers leads to their depletion. The environmental burden of the fossil fuels utilization is also very strong [1]. Many alternative methods of energy generation typically have highly varied production rates, e.g. changing from maximum generation capacity to zero within 12 hours in case of the solar energy. Such methods of energy generation become economically viable only if sufficiently efficient methods of storing the energy can be found. The existing rechargeable batteries are mostly based on ionic effects, and by their nature have severe limitations on physically implementable charging/discharging rates, as well as growing costs of production due to the use of rare and difficult to produce chemical elements in many types of more efficient batteries. On the other hand, creation of all-electronic energy storage mechanisms would provide a valuable alternative.

The only known mechanism of the energy storage based on electrons is the usual capacitor, made of two metallic plates separated by a dielectric. There are two limiting factors in such systems, namely the dielectric strength and the leakage [2,3,4,5,6], which, taken together, greatly restrict the possibility of employing capacitors as a replacement for common electrochemical energy storage systems.

Recently there has been a resurgence of interest in a potential role of electronic capacitors as energy storage devices [7,8,9,10]. Of particular interest is the possible increase of the energy density resulting from the reduction of spacing between the capacitor plates down to the nanometer scale. In such nanolayer capacitors, in which the thickness of the dielectric layer is on the order of a few nanometers, the dielectric strength was demonstrated to increase significantly [11,12,13]. Consequently, these nanocapacitors can tolerate substantially higher electric fields before a breakdown damaging of the dielectric layer would finally occur. We demonstrate here yet another important phenomenon occurring in nanolayer capacitors: The dielectric layer can be asymmetrically charged so that the amount of charge stored inside the dielectric can be many times higher than the charge accumulated on the capacitor plates. One of the most suitable, easily accessible, and actively researched materials for the dielectric layer of nanolayer capacitors is the aluminum oxide, which can be deposited using commercial atomic layer deposition machines [14,15,16]. This dielectric compound ($Al_2O_3$) typically exhibits low leakage currents, which is an important advantage in energy storage applications [13,17,18]. Since Al is a good conductor and $Al_2O_3$ is a good insulator, and both are relatively inexpensive, they can be a material of choice for building energy storage facilities. As of now, aluminum oxide finds applications in pulsed sources of power [9,10,15,19].

Here we discover that the electrons can be effectively injected into the dielectric using the field emission effect. The electrons, when injected, become trapped because the amorphous alumina has a substantial density of the electronic traps. If the temperature is sufficiently low, the injected electrons remain stable. If the temperature is increased to the point when the electrons begin to diffuse through the dielectric then the trapped charge is released to the plates of the capacitor and a battery action is observed, since the electrons mostly diffuse to the nearest capacitor plate, which is the anode. Similar effects have been considered theoretically [20,21] and now we provide an experimental evidence, obtained on capacitors with nanoscale dielectric layers. The most significant result is that the energy stored can be much greater (7.5 greater in the best case obtained so far) than the charge stored on the plates of the capacitor. The successful charge storage in the dielectric require an asymmetric electronic density distribution. If the distribution is symmetric than the amount of charge flowing to the cathode and anode would be the same. Yet, if the distribution is asymmetric then one of the electrodes will get more charges, which leads to a significant discharge current, i.e., a release of energy. We demonstrate that such asymmetric charge distributions can be created by means of the field emission.

## 2. EXPERIMENTAL DETAILS

We fabricate and study Al nanolayer capacitors of the type $Al/Al_2O_3/Al$ (samples S_1, S_2, S_3). In addition to those, for comparison, we have fabricated and studied $Cr/Al_2O_3/Cr$ capacitors (sample S_b) as well. The Al and Cr top (25 nm) and bottom (25 nm) plates were produced by thermal evaporation onto a glass substrate in a vacuum of ~$10^{-5}$ Torr. The alumina ($d_{oxide}$=10 nm for samples S_1, S_2, S_3, and 9 nm for S_b) was deposited using trimethylaluminum/$H_2O$-based ALD deposition at 80°C. The surface of the sample S_1 was $A$=1 mm$^2$, with the capacitance $C$=8 nF. For samples S_2, S_3, and S_b these parameters were $A$=2.25 mm$^2$, 2.25 mm$^2$, 1.21 mm$^2$ and $C$=12 nF, 11.43 nF, 7 nF correspondingly. Measurements at cryogenic temperatures were performed using a sample-in-vacuum dipstick (~$10^{-3}$ Torr), immersed in liquid $N_2$. To shield the samples from external electromagnetic noises the samples were placed into a Faraday cage, located inside the dipstick.

Electrical measurements were performed with Keithley 6517B electrometer. The output voltage, $V$, of this devise was applied to the capacitor through a calibrated series resistor, $R_{st}$=1 GΩ. The current in the circuit, $I$, was also measured by Keithley 6517B. The voltage on the sample (the tested capacitor), $V_S$, was computed by the formula $V_S=V-IR_{st}$. The "high" voltage terminal was always connected with the top plate of the capacitors, which means that the positive voltage corresponds to the positive potential on the top plate. The experiment was carried out in several successive phases: #1 charging the capacitor at $T$=77 K, for the time duration of $t_{ch}$, which was a few hours, at a fixed voltage on the sample $V_S$, typically a few volts; #2 discharging through a series standard resistor $R_{st}=1$ GΩ for a duration of 5 min, followed by a conformational discharging with the plates being short-circuited (without any resistor but through a copper wire) for 50 min, to further ensure that the plates of the capacitor are fully discharged; #3 warming up to room temperature, while the applied voltage was zero.

## 3. RESULTS

A typical time dependence of the charging current for a full experimental cycle is shown in Figure 1a. The cycle observed on the curve match the experimental stages outlined above, namely: #1 charging for $t_{ch}=19$ h, at $V_S=4.76$ V, #2 discharging at $T=77$ K, and #3 warming up the capacitor, while the current was measured through an ammeter connecter to the capacitor through a 1 GΩ resistor. Thus, in stage #1 we find an exponential drop of the charging current, as expected for charging of a capacitor through a resistor. In stage #2 the discharge current is shown, which is negative. The key result is the current peak observed in the stage #3, in which no voltage was applied, and the detected current was due to the extra charge trapped in the dielectric, which was released by thermal fluctuations. The time dependence of the sample temperature corresponding to all three stages is shown in Figure 1b.

In Figure 1a(insert) we show the charging current measured in the stage #1 (at temperature $T=77$ K) of the experimental cycle. Initially curve shows an exponential drop of the charging current as expected for an ideal capacitor. From the slope we find the capacitance of $C_{77K}=11.07$ nF (the corresponding charging time was $\tau=11.07$ s, while the series resistance was 1 GΩ), while a measurement performed at a higher frequency (120 Hz; performed using a LCR-meter 878A) showed a slightly lower value of 10.75 nF. We note in passing that the room temperature capacitance was about the same value, 11.43 nF, which shows high thermal stability of these nanolayer capacitors. The key fact following from the Fig.1a(insert) is that the charging current drops slow down at long times. The charging curve exhibits a «tale», which indicates that the current flowing into the capacitor is larger than the one expected for an ideal capacitor. We explain this current by the charge penetration in the dielectric, which is also known as soakage as well as dielectric absorption [17,22,23].

To better understand the charge storage in the capacitor, including its dielectric layer, we first integrate the discharge current and find the total charge exiting the capacitor. The integral charge (stage #2) was $Q_P=\int Idt \approx 51.87$ nC, which corresponded well to the charge stored on the capacitor plates estimated by the capacitor charge formula $Q_P=C_{77K}V_S=52.36$ nC. Subsequent heating of the previously discharged capacitor exhibits a much larger integral charge, which is a highly surprising result. Namely, the total charge released due to the heating of the capacitor was evaluated at $Q_D=\int Idt \approx 346$ nC, based on the time dependence plot for the discharge $I(t)$ (Figure 1(a)). Here the subscript «D» implies that this charge was stored in the dielectric and was released during the heating stage. Not that the discharge current occurs only as the temperature reaches ~225 K. An important observation here is that the direction of the heat-induced current matches the direction of the initial charging current (as measured in stage #1).

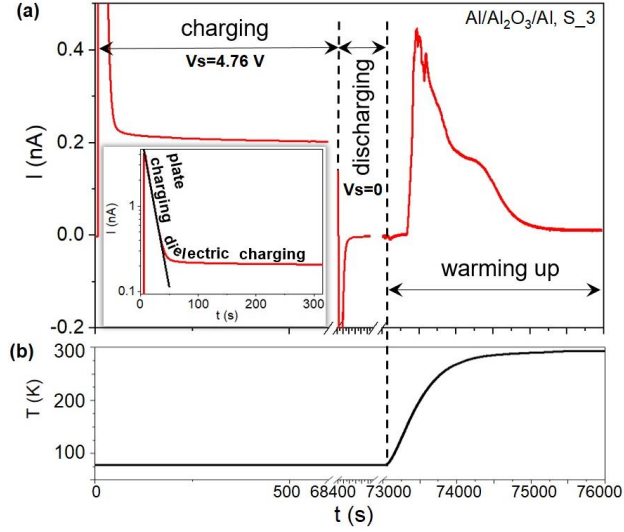

*Figure 1. Current (a) and temperature (b) as functions of time at different regimes for sample S_3: (1) charging (for 19 h at 77 K), (2) discharging for 5 min (with resistor)+50 min (with short-connected plates), at 77K, and (3) warming up to room temperature. (a, insert): The charging current as a function of time. The black line represents the best exponential fit for the charging current of the plates of the capacitor, marked as "plate charging". At times t > 40 s the measured current is larger than the extrapolated exponential dependence. The observed excess current is due to the dielectric charging current as well as the field emission leakage current (red curve marked as "dielectric charging"). (b) The temperature ramps up as a function of time. The horizontal time axis in (a) and (b) is the same.*

In Figure 2 the total charge, $Q_D$, released during the stage #3 (warming up) is shown, normalized by the calculated charge stored on the capacitor plates in the fully charged states, $Q_P=C_{77K}V_S$, where $V_S$ is the voltage on the capacitor, corresponding to the charging cycle (stage#1). The charge stored in the dielectric, $Q_D$, is obtained, as explained above, by the integration of the current flowing out of the capacitor during the warming up (stage#3). As is illustrated in Figure 2, the dielectric-stored charge depends strongly on the charging voltage. At voltages $V_S<4$ V for Al/Al$_2$O$_3$/Al and $V_S<4.7$ V for Cr/Al$_2$O$_3$/Cr the stored charge is near zero. As we increase the charging voltage to the level at which the field emission leakage current becomes significant (Figure 2 insert), we observe a strong increase of the stored charge $Q_D$. The normalized charge reaches its maximum value of $Q_D/Q_P=6.6$, corresponding to the charging voltage $V_S=4.76$ V for Al/Al$_2$O$_3$/Al. Thus, we discover that the charge stored in the dielectric can be much larger than the charge stored on the capacitor plates. The results on a Cr film capacitor confirmed the finding: The maximum normalized charge was $Q_D/Q_P=6.9$ (corresponding to the charging voltage $V_S=5.1$ V). Further increase of the charging voltage leads to a steep drop of the stored charge $Q_D$.

The voltage-current dependence of two typical nanolayer capacitors is shown in Figure 2(insert). At low voltages, the leakage current is so low that it is undetectable. As the voltage exceeds some threshold, $V_{th}$, we measure a leakage current which increases roughly exponentially with the voltage on the capacitor, $V_S$. The V-I curves show some hysteresis but only during the first cycle of the sweeping voltage. If the voltage is increased to the maximum and decreased to zero and then the V-I curve is measured again then the hysteresis does not occur. We explain this fact by the conjecture that the electrons enter the dielectric in the first charge cycle, get trapped by the dielectric and do not move away as the voltage is reduced to zero. The circles on the Figure 2(insert) correspond to a V-I curve measured after the sample was warmed up to the room temperature and then cooled again to 77 K. This curve shows that the effect is reproducible, and the charges can indeed be removed from the dielectric if the temperature is increased to room temperature.

An important observation, which we want to emphasize here, is that if the charging voltage of the capacitor is chosen such that it is less than $V_{th}$ then the stored charge (see Figure 2) is near zero. If the charging current is slightly larger than $V_{th}$, then the stored charge shows a maximum. If the charging voltage is significantly larger than $V_{th}$, then the stored charge drops quickly (Figure 2). For example, for the Al capacitor $V_{th}=4$ V and the maximum of the stored charge is observed at $V_S=4.76$ V. Yet, if the charging voltage is chosen $V_S=4$ V or $V_S=5$ V then the stored charge is near zero. For the chromium-plate capacitor the results are similar. The threshold voltage is $V_{th}=4.7$ V. The maximum of the stored charge is observed at $V_S=5.1$ V. Yet, if the charge voltage is chosen, e.g., $V_S=5$ V or $V_S=5.3$ V then the stored charge is more than two times lower.

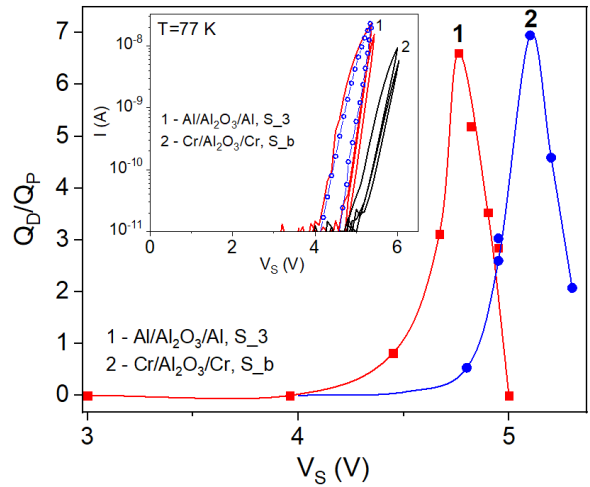

*Figure 2. Normalized charge stored in the dielectric of nanolayer capacitors. The blue curve represent the Cr based capacitor and the red curve – Al based capacitor. The horizontal axis gives the voltage at which the capacitors were charged (stage #1). The stored charge, $Q_D$, was measured in stage #3, in which no voltage was applied and the released current (measured using Keithley 6517B), was integrated to give the total stored charge in the dielectric. (Insert) The insert shows the V-I curves for the two types of capacitors, Al based, and Cr based. The hysteresis is seen only in the first sweep of the voltage. The blue circles represent a curve measured after the sample was warmed up and cooled down again, to show reproducibility.*

Penetration of charges into the dielectric layer occurs by means of quantum tunneling, namely the field emission. Quantum tunneling is a process which conserves energy so the charges tunnel to the regions of the dielectric which are near the anode. This is because the energy of the electronic traps located near the anode becomes equal to the Fermi energy of the electrons in the cathode, assuming the bias voltage is optimally tuned to maximize the charge storage effect.

It is also known that the tunneling phenomenon is an exponentially slow phenomenon. Thus, performing the charging procedure longer can help to fill more charge traps. Indeed, our experiments showed that the trapped charge $Q_D$ can be increased if the charging time is increased. This is illustrated in Figure 3. There we present normalized trapped charge (charge stored in the dielectric) versus the time over which the voltage was applied during the charging process. There is an increased of $Q_D$ as the time is increased up to ~10-20 hours. In this respect such nanolayer capacitors resemble rechargeable batteries which require a long time to be charged.

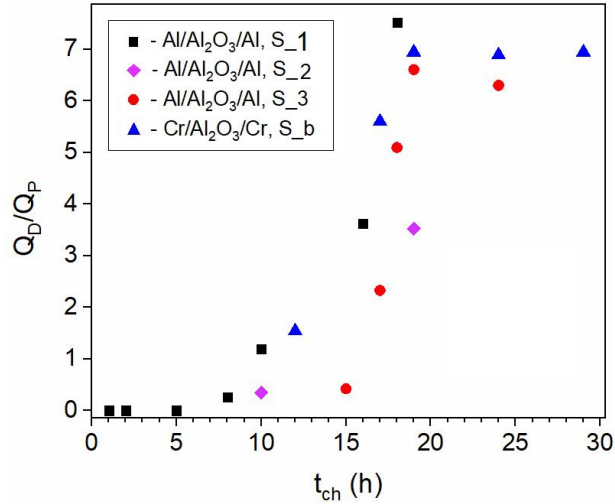

*Figure 3. Total charge, stored in the dielectric and released during the warming up phase (stage #3), plotted versus the charging time. The charging is performed at a fixed voltage, which was 4.82 V, 4.82 V, 4.76 V and 5.1 V for samples S_1, S_2, S_3, and S_b correspondingly. The charge stored in the dielectric is normalized (divided) by the charge stored on the capacitor plates. It is clear that the charge stored in the dielectric nanolayer of the capacitor can be up to seven times higher than the charge stored on the capacitor plates during the charging process.*

We also performed experiments to establish stability of the stored charge. The sample S_3 was charged at a voltage $V_S$=4.82 V, discharged for 5 min with the resistor and discharged for 50 min with short-circuited plates, at 77 K. Then the stored charge was measured giving us $Q_D$=275 nC. In the subsequent experiment all steps were the same but the discharge with shorted plates continued for 136 h. The resulting charge was $Q_D$ =214 nC. Thus, we demonstrate that the charge is quite stable even if the plates are short-circuited for a long time.

The maximum efficiency observed in our experiments was 7.5, in the sense that the charge stored in the dielectric was 7.5 times higher than the charge stored on the capacitor plates during the charging. To our knowledge this is much larger than any value reported previously [24,25,26]. We also estimated the energy density. The total energy was $W_D=Q_D^2/2C \approx 5.2$ μJ. The volume of the dielectric and the mass of the dielectric have been evaluated using standard table value. Then the resulting energy density is $w=W_D/(Ad_{oxide})$=520 J·cm$^{-3}$. The results are shown in Figure 4. An important finding there is that, to achieve high energy density, the charging voltage needs to be precisely tuned and the charging time needs to be sufficiently long. This is higher than ~307 J·cm$^{-3}$, which is the highest value we are aware of, achieved previously in Au/SrTiO$_3$/La$_{0.67}$Sr$_{0.33}$MnO$_3$ capacitors [27]. If expressed per unit mass the energy density is ~200 J·g$^{-1}$, assuming the density [28] of the alumina 2.6 g·cm$^{-3}$.

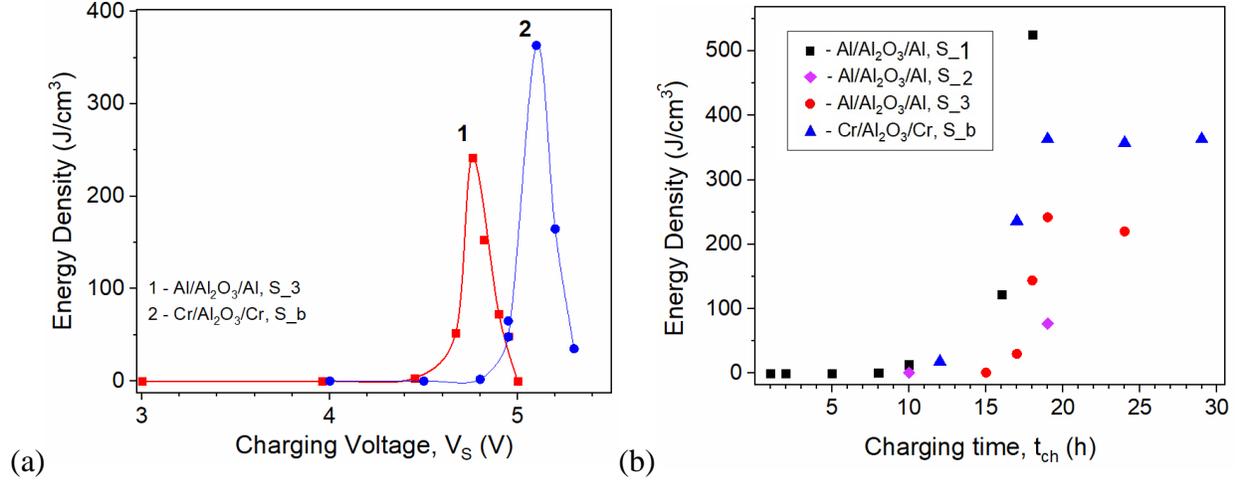

*Figure 4. (a) Energy density, stored in the nanolayer dielectric, versus the voltage applied to the capacitor during the charging process. The results show that the function is sharply peaked. The peak corresponds to such voltage at which the electrons can tunnel into the dielectric layer, but only in the region near the anode. (b) The energy density as a function of the charging time. The soaking of the dielectric with electrons appears to saturate after 10-20 hours.*

## 4. DISCUSSION

The observed current produced upon the heating of the capacitor is entirely due to release of the charge previously accumulated in the dielectric layer [22,23,29]. Furthermore, the direction of this additional current demonstrates that it cannot be attributed to dielectric polarization from hindered movement of dipoles, which constitutes one of the two main mechanisms of dielectric response [30]. A discharge current from this type of polarization flows in the same direction as the conventional discharge current from the capacitor plates, i.e. in the opposite direction of the charging current. Yet, the discharge current we measure as we increase the temperature flows in the same direction as the charging current. We suggest that the secondary discharge current results from the other type of dielectric response, involving the charge accumulation due to tunneling and subsequent trapping of electrons at localized sites in the dielectric layer. This process leads to the formation of localized trapped charge in the dielectric, which cannot be dissipated by thermal fluctuations at cryogenic temperatures even when the applied voltage is reduced to zero.

The main factor determining the filling of localized electron traps appears to be the charging voltage. At low charging voltages $V<V_{th}$ there is no leakage current through the capacitor, and the electron traps in the dielectric layer primarily remain unoccupied (Figure 5(a)). Here, $V_{th}$ is the threshold voltage at which the field emission leakage current through the capacitor begins (Figure 2, insert). The filling of traps starts at $V>V_{th}$, at the onset of the field emission. Note that the average energy of the charge traps located near the anode is lower since the electrons are attracted to the anode. At charging voltages exceeding $V_{th}$ electrons can tunnel into localized trap sites with the energy below the Fermi level, i.e., only those which are located close to the anode [31,32] (Figure 5(b)). Thus, the field emission effect produces a strongly nonuniform distribution: More charges are injected into the region near the anode while the region near the cathode is completely depleted of extra electrons since the electrons cannot tunnel in the traps in that region, since the energy of such traps is higher. The resulting trapped charge distribution in the dielectric layer is substantially inhomogeneous, with a progressively larger concentration of trapped electrons closer to the anode.

If the capacitor is discharged at low temperature, the majority of trapped electrons remain in the dielectric layer, since the thermal fluctuations are insufficient for the electrons to escape from the trapping sites. This is clear also from the fact that at low temperature the measured diffusive electronic conductivity is undetectable (i.e., the conductivity occurring at low bias).

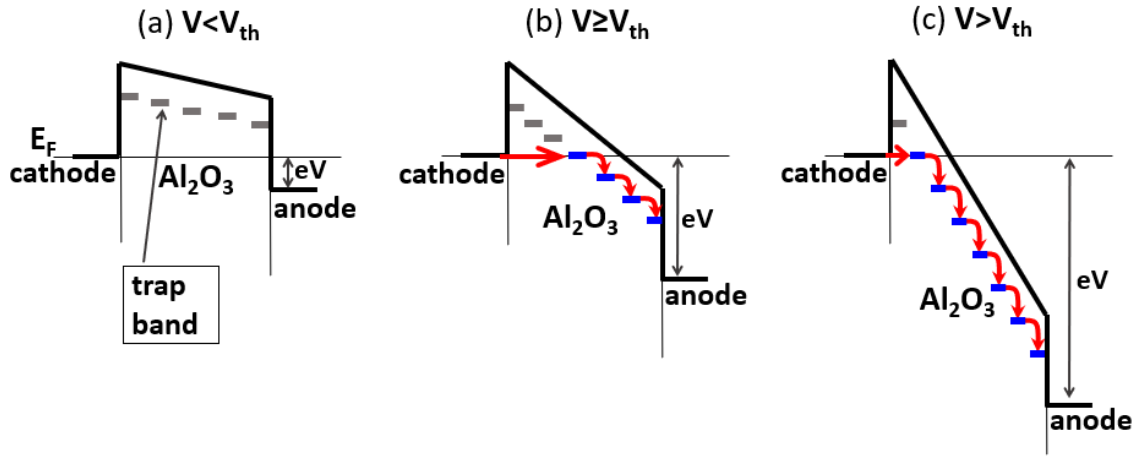

*Figure 5. Schematic representation of the suggested mechanism: Charge injection into the dielectric region adjacent to the anode, by the field emission mechanism. The grey lines represent empty traps, and the blue lines represent the filled traps. Electrons can only fill up the traps which are lower in energy compared to the Fermi level.*

When the temperature is increased in the later stage of the experiment, the energy of thermal fluctuations becomes sufficient for freeing electrons from the trapping sites. Therefore, this stage of the experiment essentially represents a thermal spectroscopy analysis (see Figure 6) of electron traps: At any given temperature the electrons leaving the dielectric layer are such that are trapped at sites with the trapping energy of the order of $k_BT$. Therefore, at a sufficiently high temperature many electrons regain mobility and move towards the nearest electrode. Since the concentration of trapped charges was larger close to the anode, we observe the asymmetric discharge current, matching the direction of the charging current (Figure 1(a)). In this thermal escape process the electrons diffuse to the anode since it is the closest electrode to the majority of them.

If the charging voltage greatly exceeds $V_{th}$, electrons tunneling from the cathode can access traps with higher energy (Figure 5(c)), i.e., those near the cathode. This leads to a more homogenous charge distribution in the dielectric layer. As a result, the difference between the numbers of electrons freed from traps closer to the anode and from those closer to the cathode upon heating of the capacitor is reduced (Figure 2). For a certain value of the charging voltage, the charge reaching both electrodes upon the thermal discharge becomes identical, and we no longer observe a discharge current.

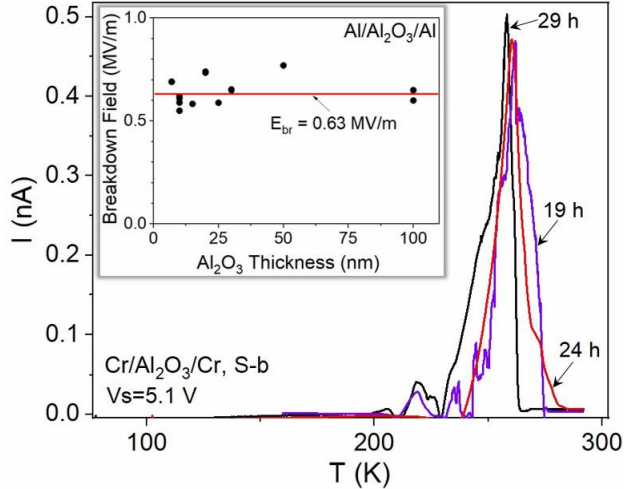

*Figure 6. Discharge current versus temperature, at zero applied voltage. The plots show the thermal escape of electrons from their traps. The curves represent a form of thermal spectroscopy. The time (19 h, 24 h and 29 h) indicated is the charging time, while the charging voltage was $V_S$=5.1 V. (Insert). The breakdown electric field (i.e., the dielectric strength) as a function of the dielectric film thickness, obtained on capacitors with Al electrodes. The dielectric strength appears independent of the film thickness at the scale between 7 nm and 100 nm.*

It should be noted that the efficiency of a capacitor to store energy is also limited by its dielectric strength. Yet, we have observed (Figure 6, insert) that the strength of the electric field at which the breakdown occurs is considerably larger than the field needed to charge the dielectric and it is independent of the thickness (Figure 6, insert) of the dielectric. Thus, employing nanolayer dielectrics, needed for an efficient field emission charging, does not negatively impact the capacitor performance.

## 5. CONCLUSIONS

We fabricate and study $Al/Al_2O_3/Al$ and $Cr/Al_2O_3/Cr$ nanolayer capacitors to optimize the process of the energy storage by purely electronic mechanisms (without involvement of ionic effects). It is found that the field emission process is able to selectively charge the layer of the dielectric close to the anode, which can be referred to as spatially selective dielectric soakage. The charge can be reliable stored in the dielectric layer and remains intact even if the capacitor plates are short-circuited, provided that the temperature is sufficiently low, i.e., lower than ~200 K in the considered cases. The charge can be recovered by heating the system to room temperature, through the effect of the dielectric relaxation. We discover that the charge stored in the dielectric exceeds strongly the charge stored on the plates of the capacitor. The largest ration observed, between the charge stored on the dielectric and the charge stored on the capacitor plates, was ~7.5. The total charge stored in the dielectric corresponds to the energy density ~520 J·cm$^{-3}$. Also, the nanolayer capacitors show excellent thermal stability as their capacitance does not change with temperature.

### ACKNOWLEDGMENTS
This work was supported by the Air Force grant AF FA9453-18-1-0004.

# REFERENCES

[1] Chu S, Majumdar A 2012 Opportunities and challenges for a sustainable energy future *Nature* **488** 294-303.

[2] Chiang T-H and Wage J F 2018 Electronic Conduction Mechanisms in Insulators *IEEE Transactions on Electron Devices* **65(1)** 223-230.

[3] Chiu F-C, Lee C-Y, and Pan T-M 2009 Current conduction mechanisms in $Pr_2O_3$/oxynitride laminated gate dielectrics *J. Appl. Phys*. **105** 074103.

[4] Lin H C, Ye P D, and Wilk G D 2005 Leakage current and breakdown electric field studies on ultrathin atomic-layer deposited $Al_2O_3$ on GaAs *Appl. Phys. Lett*. **87** 182904.

[5] Groner M D, Elam J W, Fabreguette F H, George S M 2002 Electrical characterization of thin $Al_2O_3$ films grown by atomic layer deposition on silicon and various metal substrates *Thin Solid Films* **413**, 186-197.

[6] Koda Y, Sugita H, Suwa T, Kuroda R, Goto T, Teramoto A, and Sugawa S 2016 Low Leakage Current $Al_2O_3$ Metal-Insulator-Metal Capacitors Formed by Atomic Layer Deposition at Optimized Process Temperature and $O_2$ Post Deposition Annealing *ECS Transactions* **72(4)** 91-100.

[7] Zhao X, Sanchez B M, Dobson P J, and Grant P S 2011 The role of nanomaterials in redox-based supercapacitors for next generation energy storage devices *Nanoscale* **3**, 839.

[8] Belkin A, Bezryadin A, Hendren L, and Hübler A W 2017 Recovery of Alumina Nanocapacitors after High Voltage Breakdown *Sci Rep.* **7** 932.

[9] Yao K, Chen S, Rahimabad M, Mirshekarloo M S, Yu S H, Tay F E H, Sritharan T, Lu L 2011 Nonlinear dielectric thin films for high-power electric storage with energy density comparable with electrochemical supercapacitors *IEEE Trans. Son. Ultrason.* **58** 1968-1974.

[10] Correia T M, McMillen M, Rokosz M K, Weaver P M, Gregg J M, Viola G, Cain M G 2013 A lead-free and high-energy density ceramic for energy storage applications *J. Am. Ceram. Soc*. **96** 2699-2702.

[11] Sirea C, Blonkowski S, Gordonb M J, and Baron T 2007 Statistics of electrical breakdown field in $HfO_2$ and $SiO_2$ films from millimeter to nanometer length scales *Appl. Phys. Lett*. **91** 242905.

[12] Lyon D and Hübler A 2013 Gap size dependence of the dielectric strength in nano vacuum gaps *IEEE Transactions on Dielectrics and Electrical Insulation* **20(4)** 1467-1471.

[13] Spahr H, Reinker J, Bülow T, Nanova D, Johannes H-H, and Kowalsky W 2013 Regimes of leakage current in ALD-processed $Al_2O_3$ thin-film layers *J. Phys. D: Appl. Phys*. **46** 155302.

[14] Etinger-Geller Y, Zoubenko E, Baskin M, Kornblum L, and Pokroy B 2019 Thickness dependence of the physical properties of atomic-layer deposited $Al_2O_3$ *J. Appl. Phys*. **125** 185302.

[15] Pan Z, Ding Q, Yao L, Huang S, Xing S, Liu J, Chen J, and Zhai J 2019 Simultaneously enhanced discharge energy density and efficiency in nanocomposite film capacitors utilizing two-dimensional $NaNbO_3@Al_2O_3$ platelets *Nanoscale* **11** 10546-10554.

[16] Attiq ur-rehman, Ashraf M W, Shaikh H, Alhamidi A, Ramay S M, Saleem M 2020 Yttrium incorporated $BiFeO_3$ nanostructures growth on two step anodized $Al_2O_3$ porous template for energy storage applications *Ceramics International* **46** 7681-7686.

[17] Hillmann S, Rachut K, Bayer T J M, Li S, and Klein A 2015 Application of atomic layer deposited $Al_2O_3$ as charge injection layer for high-permittivity dielectrics *Semicond. Sci. Technol*. **30** 024012.

[18] Ding S-J, Hu H, Lim H F, Kim S J, Yu X F, Zhu C, Li M F, Cho B J, Chan D S H, Rustagi S C, Yu M B, Chin A, Kwong D-L 2003 High-Performance MIM Capacitor Using ALD High-κ $HfO_2$-$Al_2O_3$ Laminate Dielectrics *IEEE Electron Device Letters* **24(12)** 730-732.

[19] Winsor P, Scholz T, Hudis M, Slenes K M 1999 Pulse power capability of high energy density capacitors based on a new dielectric material *Proc. IEEE Int. Pulsed Power Conf.* **1**, 102-105.

[20] Wintle H J 1973 Absorption current, dielectric constant, and dielectric loss by the tunnelling mechanism *Journal of Applied Physics* **44** 2514.

[21] Lowell J 1979 Tunnelling between metals and insulators and its role in contact electrification *J. Phys. D: Appl. Phys.* **12** 1541-1554.

[22] Bezryadin A, Belkin A, Ilin E, Pak M, Colla E V, and Hübler A 2017 Large energy storage efficiency of the dielectric layer of graphene nanocapacitors *Nanotechnology* **28**, 495401.

[23] Ilin E, Burkova I, Draher T, Colla E V, Hübler A, and Bezryadin A 2020 Coulomb barrier creation by means of electronic field emission in nanolayer capacitors *Nanoscale* **12**, 18761-18770.

[24] Gupta H, Singh K, and John T 2014 Analysis of Dielectric Absorption in Capacitors *Journal of Academia and Industrial Research* **3(6)** 255-257.

[25] Teverovsky A 2014 Absorption Voltages and Insulation Resistance in Ceramic Capacitors with Cracks *IEEE Transactions on Dielectrics and Electrical Insulation* **21(5)** 2020-2027.

[26] Kuenen J C and Meijer G C M 1996 Measurement of dielectric absorption of capacitors and analysis of its effects on VCOs *IEEE Trans. Instrum. Measur.* **45(1)** 89-97.

[27] Hou C, Huang W, Zhao W, Zhang D, Yin Y, and Li X 2017 Ultrahigh Energy Density in $SrTiO_3$ Film Capacitors *ACS Applied Materials & Interfaces* **9(24)** 20484-20490.

[28] Groner M D, Fabreguette F H, Elam J W, George S M 2004 Low-Temperature $Al_2O_3$ Atomic Layer Deposition *Chem. Mater.* **16(4)** 639-645.

[29] Belkin A, Ilin E, Burkova I, and Bezryadin A 2019 Reversed Photoeffect in Transparent Graphene Nanocapacitors *ACS Appl. Electron. Mater.* **1** 2671-2677.

[30] Jonscher A 1999 Dielectric relaxation in solids *J. Phys. D: Appl. Phys*. **32** R57.

[31] Jegert G, Kersch A, Weinreich W, Schröder U, and Lugli P 2010 Modeling of leakage currents in high-κ dielectrics: Three-dimensional approach via kinetic Monte Carlo *Appl. Phys. Lett*. **96** 062113.

[32] Weiler B, Haeberle T, Gagliardi A, and Lugli P 2016 Kinetic Monte Carlo of transport processes in Al/$AlO_x$/Au-layers: Impact of defects *AIP Adv*. **6** 095112.